\documentclass[12pt]{article}
\pdfoutput=1 
\usepackage[top    = 1.1in, bottom = 1.1in, left   = 1in, right  = 1in]{geometry}

\renewcommand{\baselinestretch}{1}


\usepackage{amsmath}
\usepackage{graphicx,psfrag,epsf}
\usepackage{enumerate}
\usepackage{natbib}
\usepackage{amssymb}

\date{\today}

\newcommand{\blind}{1}
\newcommand{\papertitle}{Bayesian analysis of $^{210}Pb$ dating} 

\begin{document}


	\def\spacingset#1{\renewcommand{\baselinestretch}%
		{#1}\small\normalsize} \spacingset{1}

	
	\if1\blind
	{
		\title{\textbf{\papertitle}}
		
		\author{Marco A Aquino-L\'opez\thanks{
				School of Natural and Built Environment,
				Queen's University Belfast,
				Belfast, BT7-1NN. 
				email: \texttt{maquinolopez01@qub.ac.uk}, \texttt{maarten.blaauw@qub.ac.uk}} \thanks{Corresponding author.}  \and
			Maarten Blaauw$^*$ \and
			J Andr\'es Christen\thanks{
				Centro de Investigaci\'on en Matem\'aticas (CIMAT),
				Jalisco s/n, Valenciana, 36023 Guanajuato, GT, Mexico.
				email: \texttt{jac@cimat.mx}  } \and
			Nicole K. Sanderson\thanks{
				College of Life and Environmental Sciences, University of Exeter, 
				Exeter, EX4-4QJ.
				email: \texttt{N.K.Sanderson@exeter.ac.uk}} }
		
		\maketitle
	} \fi
	
	\if0\blind
	{
		\bigskip
		\bigskip
		\bigskip
		\begin{center}
			{\LARGE\bf \papertitle}
		\end{center}
		\medskip
	} \fi
	
	\bigskip
\begin{abstract}
	In many studies of environmental change of the past few centuries, $^{210}Pb$ dating is used to obtain chronologies for sedimentary sequences. One of the most commonly used approaches to estimate the ages of depths in a sequence is to assume a constant rate of supply (CRS) or influx of `unsupported' $^{210}Pb$ from the atmosphere, together with a constant or varying amount of `supported' $^{210}Pb$. Current $^{210}Pb$ dating models do not use a proper statistical framework and thus provide poor estimates of errors. Here we develop a new model for $^{210}Pb$ dating, where both ages and values of supported and unsupported $^{210}Pb$ form part of the parameters. We apply our model to a case study from Canada as well as to some simulated examples. Our model can extend beyond the current CRS approach, deal with asymmetric errors and mix $^{210}Pb$ with other types of dating, thus obtaining more robust, realistic and statistically better defined estimates.   
\end{abstract}	
	
	\noindent%
	{\it Keywords:} $^{210}Pb$ dating, Chronology, Bayesian Analysis, MCMC, Sediment core.
	\vfill
	
	\newpage
	\spacingset{1.45} 

\section{Introduction}

Radiometric dating is a series of techniques used to date material containing radioactive elements \citep{Gunten1995} which decay over time. These techniques use the radioactive decay equation ($N(t)=N_0e^{-\lambda t}$, where $N(t)$ is the quantity of a radioactive element left in the sample at time $t$, $N_0$ is the initial quantity, and $\lambda$ is the element's decay constant) to infer the age of the material being investigated.

$^{210}Pb$ (lead-210) is a radioactive isotope which forms part of the $^{238}U$ (uranium) series. $^{238}U$ (solid) is contained within most rocks and decays into $^{226}Ra$ (radium, solid), which later decays into $^{222}Rn$ (radon, gas). Since $^{222}Rn$ is a gas, a proportion escapes to the atmosphere where it decays into $^{210}Pb$ (solid) which is later transported to the earth's surface by precipitation. $^{210}Pb$ deposited this way is labelled ``unsupported'' or excess $^{210}Pb$ ($P^U$). On the other hand, $^{222}Rn$ which decays in situ becomes what is labelled ``supported'' $^{210}Pb$ ($P^S$). By distinguishing between supported and unsupported $^{210}Pb$ one can determine the age of the sediment through measuring the $^{210}Pb$ at a depth $d$ and compare it to the rest of the sediment.

The sediments within lakes, oceans and bogs contain layers of biotic and abiotic fossils, which can be used as indirect time-series of environmental dynamics as the sediments accumulate over time.. Whereas both unsupported and supported $^{210}Pb$ decay over time, supported $^{210}Pb$ is replenished through decay from radon contained within the sediment. That is why the concentration of supported $^{210}Pb$ remains at equilibrium while that of the unsupported $^{210}Pb$ decreases and eventually reaches zero. This is the basis for $^{210}Pb$ dating. Given its relatively short half-life of 22.3 years, $^{210}Pb$ has been used to date recent ($< 200$ yr old) sediments. This time period is of particular importance when studying the effects of humans on the environment.
Unfortunately, the current dating models were not created within a statistical framework. This has given place to a series of uncertainty approximations \citep{ Appleby2001, Sanchez-Cabeza2014}. Here we introduce both a new treatment of $^{210}Pb$ data and a dating model created within a statistical framework, with the objective of providing more reliable measures of uncertainty.

\section{Modelling of $^{210}Pb$ data}\label{sec:210mode}

As outlined above, within sediment $^{210}Pb$ is naturally formed from two sources -- from surrounding sediment and rocks containing $^{238}U$ (supported), and from the atmosphere through $^{220}Rn$ (unsupported). Modelling these two sources of $^{210}Pb$ is crucial to the development of age-depth models. Since supported and unsupported $^{210}Pb$ are indistinguishable from each other, in order to model both sources, we have to make assumptions depending on the measurement techniques used.  Measurements of $^{210}Pb$ can be obtained by alpha or gamma spectrometry. The latter technique also provides estimates of other isotopes such as $^{226}Ra$, which can be used as a proxy of the supported $^{210}Pb$ in a sample \citep{Krishnaswamy1971}.

\subsection{Supported $^{210}Pb$}

If gamma spectrometry is used, supported $^{210}Pb$ can be assumed to be equal to the concentrations of $^{226}Ra$. When the sediments are analysed using alpha spectrometry, $^{226}Ra$ measurements are not available and estimates of the supported activity can only be obtained by analysing sediment which reached background (samples which no longer contain unsupported $^{210}Pb$). When alpha spectrometry is used, a constant supported $^{210}Pb$ is assumed. These two different ways of inferring the supported activity can be formalised by the following equations: 
\begin{eqnarray}
P_i^{T} =&  P_i^{S}+P_{i}^{U}, \label{eq:S_Ra}\\
P_i^{T} =&  P^{S}+P_{i}^{U},	\label{eq:S_Pb}
\end{eqnarray}
where $P_i^{T}$ is the total $^{210}Pb$, $P_i^{U}$ is the unsupported $^{210}Pb$ and $P_i^{S}$ is the supported $^{210}Pb$ in sample $i$. Depending on the site and availability of measuring techniques, one of these equations can be used to differentiate supported from unsupported $^{210}Pb$.

\subsection{Unsupported $^{210}Pb$}

In order to model the unsupported $^{210}Pb$, some assumptions have to be made regarding the precipitation of this material from the atmosphere.  A reasonable assumption for this phenomenon is the constant flux or rate of supply \citep{Appleby1978}, which implies that for fixed periods of time the same amount of $^{210}Pb$ is supplied to the site.

Following \cite{Appleby1978}, the assumption of a constant rate of supply implies that the initial concentration of $^{210}Pb$ at depth $x$ (which is linked to age by a function $t(x)$), $P^U_0(t(x))$, weighed by the dry mass sedimentation rate $r(t(x))$, is constant throughout the core: 
\begin{eqnarray}
P^U_0(t(x))r(t(x))=\Phi \label{eq:const}
\end{eqnarray}
where $\Phi$ is a constant. The dry sedimentation rate is the speed at which the sediment accumulates, weighed by the sediment's density at such depth, i.e.
\begin{eqnarray}
r(t(x)) =\rho(x)\frac{dx(t)}{dt}  , \label{eq:drysed1}
\end{eqnarray}
where $\rho(x)$ is defined as the density of the sediment at depth $x$ and $\frac{dx(t)}{dt}$ is the rate at which the core accumulates with respect to time. Considering that the relationship between depth and time is expressed by the function $t(x)$, then $x(t)$ is the inverse function of time, and since $t(x)$ is a one-to-one function 
\begin{eqnarray}
r(t(x)) =\rho(x)\left[\frac{dt(x)}{dx}\right]^{-1}  . \label{eq:drysed2}
\end{eqnarray}

Since $^{210}Pb$ is a radioactive isotope it follows from the radioactive decay equation that 
\begin{eqnarray}
P^U(x)=P^U_0(t(x))e^{- \lambda t(x) }, \label{eq:decayeq1}
\end{eqnarray}
where $P^U(x)$ is the concentration of unsupported $^{210}Pb$ found at depth $x$, and $\lambda$ is the $^{210}Pb$ half-life. Using equations (\ref{eq:const}), (\ref{eq:drysed2}), and (\ref{eq:decayeq1}) the following relationship is obtained,
\begin{eqnarray}
\rho(x)P^U(x)= \frac{dt(x)}{dx} ~ \Phi e^{- \lambda t(x) } . \label{eq:coreequa1}
\end{eqnarray}
Considering that $^{210}Pb$ is measured over a slice or section of the sediment, this relationship has to be integrated over such section to be related to the corresponding measurement, that is,
\begin{eqnarray}
A^U_{(a,b)}&=\int_{a}^{b}\rho(z)P^U(z)dz\\  \nonumber \label{eq:sampleeq} 
&=\int_{a}^{b} \Phi e^{- \lambda t(z) }\frac{dt(z)}{dz} dz \\ 
& =\int_{t(a)}^{t(b)} \Phi e^{- \lambda y } dy,  
\end{eqnarray}
where $(a,b)$ are the lower and upper depths of the sample respectively and $A^U_{(a,b)}$ is the activity in section $(a,b)$. Equation (\ref{eq:sampleeq}) provides a link between the age-depth function $t(x)$ and the unsupported activity in a given section. This is the primary tool to construct an age-depth model based on a constant rate of supply.

\section{Current approach}

The Constant Rate of Supply (CRS) \citep{Goldberg1963,Robbins1978,Appleby1978} model is the most commonly used $^{210}Pb$ dating model. It uses the constant rate of supply assumption presented in section \ref{sec:210mode}, and the following equations to obtain a chronology: 
\begin{eqnarray}
A^U(x)&=\int_x ^\infty \rho(z) P^U(z) dz, \label{eq:A(x)}\\
A^U(0)&=\int_{0}^{\infty} \rho(z)P^U(z)  dz, \label{eq:A(0)}\\ 
t(x)&=\frac{1}{\lambda}\ln\left(\frac{A(0)}{A(x)} \right), \label{eq:txCRS}	    
\end{eqnarray}
where $A^U(x)$ is the remaining unsupported activity below $x$, and $A^U(0)$ is the unsupported activity in the whole core. The CRS model can be summarized by equation (\ref{eq:txCRS}) and from its term $A^U(0)$ one can deduce that this model depends strongly on measuring activity throughout the whole core. The effect of wrongly estimating this variable is described in \cite{Appleby1998}. If the activity cannot be measured throughout the entire core, interpolation is suggested \citep{Appleby2001}. Moreover, if the lowest sample has not reached background, and thus still contains unsupported $^{210}Pb$, extrapolation is suggested. 

Because this model is based only on the unsupported activity, precise estimates of supported $^{210}Pb$ are necessary in order to obtain reliable estimates of the unsupported $^{210}Pb$. Depending on the equipment used to obtain the $^{210}Pb$ concentrations, and on the model used to distinguish supported from unsupported $^{210}Pb$, this could be problematic. Wrongly estimating this variable will directly impact the estimate of $A(0)$ which will in consequence affect the resulting chronology.

\subsection{Example}\label{sec:HP1CCRS}

To show the results of the current approach and later compare them to ours, data obtained from a site in Havre-St-Pierre, Quebec, Canada will be used. The core (HP1C) was obtained in July 2012 and was analysed using alpha spectrometry at Exeter University, UK . Table \ref{Tab:DataHP1C} contains the data from core HP1C. As previously mentioned, alpha spectrometry does not provide estimates of $^{226}Ra$ as is the case for beta spectrometry, but instead, contrary to the latter, it can measure far smaller quantities of $^{210}Pb$. To date this core, the CRS model was calculated using the recommendations in \cite{Appleby2001}.  

\begin{figure}[ht!]
	\centering
	\includegraphics[height=.9\linewidth]{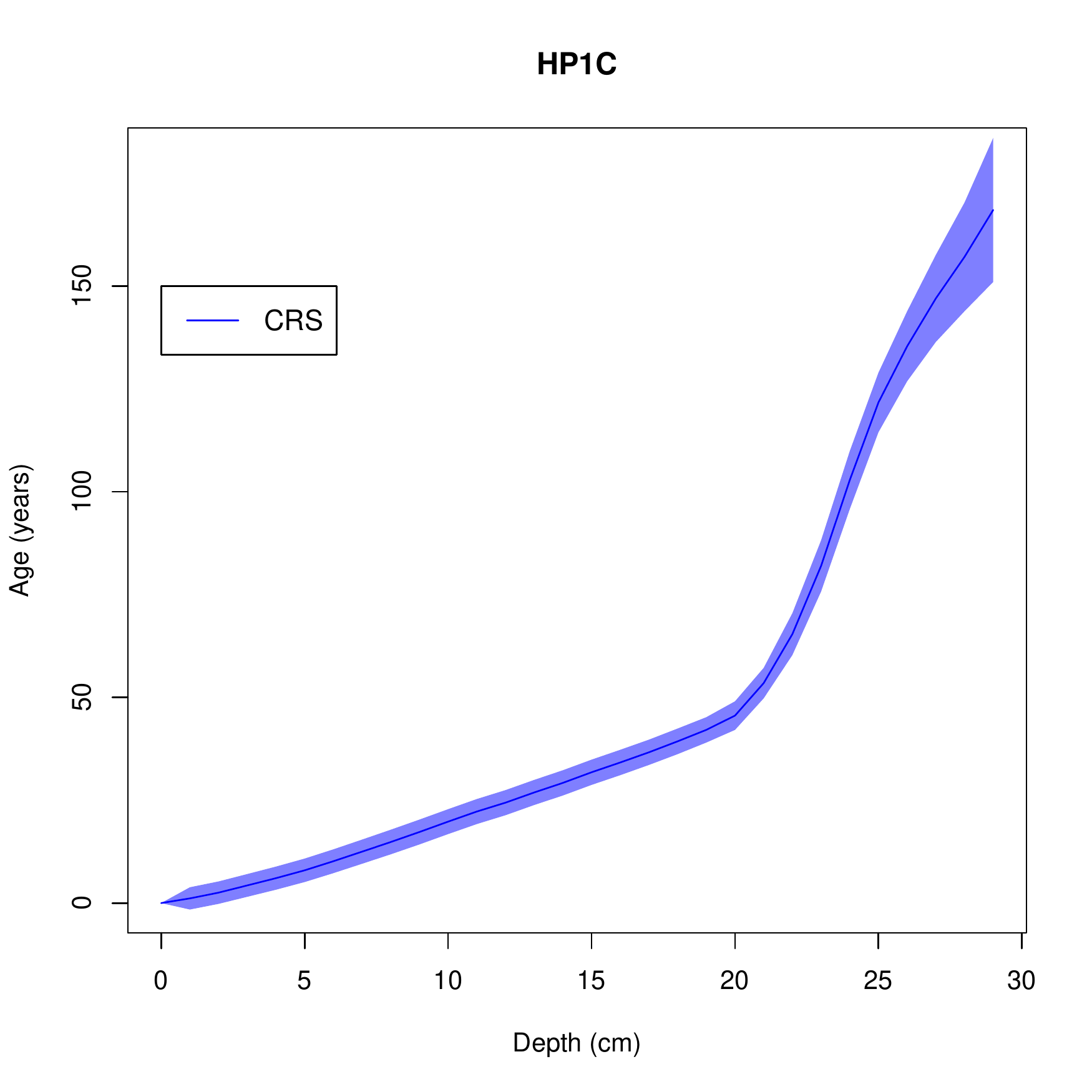}
	\caption{Dating of HP1C obtained by the CRS model \protect\citep{Appleby2001} showing the mean and 95\% confidence intervals.}
	\label{fig:CRS}
\end{figure}

One of the first steps to apply the CRS model is to identify the supported $^{210}Pb$. For this purpose the last 4 samples were averaged to obtain an estimate of $8.11 \frac{Bq}{kg}$ and a standard deviation of $1.01$ for the supported activity. This value was subtracted from the total $^{210}Pb$ for each sample, to obtain estimates of unsupported activity. Following \cite{Appleby2001} one can obtain the dating shown in Figure \ref{fig:CRS}. This methodology requires very strong assumptions regarding independence, given the fact that it uses accumulated activity as the primary tool for inference. We now introduce our formal statistical approach for $^{210}Pb$ dating, to solve this and several other issues inherent in the usual CRS technique just described.

\section{A statistical approach to $^{210}Pb$ dating}\label{chap:modelBay}

Let the concentration of $^{210}Pb$ in a sample taken from section $(x_i-\delta,x_i)$ be a random variable $p_i$. In this case, it is important to clarify that this is not the cumulative concentration or activity, from the surface to depth $x_i$, but rather it is the concentration found from depth $x_i-\delta$ to $x_i$ where $\delta$ is the sample's thickness. Each concentration of $^{210}Pb$ ($p_i$) is measured separately and therefore it is safe to assume that each sample is independent of the other measurements and is normally distributed with mean the true concentration $P_i$, and variance as reported by the laboratory:

\begin{eqnarray}
p_i\mid P_i\sim \mathcal{N} \left( P_i, \sigma_i^2 \right). 
\end{eqnarray}

As mentioned above, the supported $^{210}Pb$ is assumed to be in equilibrium throughout the core,
which means that it remains constant through all depths. If $^{236}Ra$ measurements are available, a supported $^{210}Pb$ value per sample can easily be included by letting $P_i^{S}$ be different at each depth. It is important to note that this will greatly increase the number of parameters, and should only be used when the hypothesis of a constant supported concentration has been shown to be unreasonable. If a constant supported $^{210}Pb$ is valid, then we can use equation (\ref{eq:S_Pb}) to infer the supported $^{210}Pb$.

Now, assuming a constant rate of supply, as described in section \ref{sec:210mode}, for the unsupported $^{210}Pb$, the activity in sample $i$ can be obtained as follows 
\begin{eqnarray}
A_{i}^{U} &=&	\int_{x_{i}-\delta}^{x_i} \rho_i(z) P_i^{U}(z)dz \nonumber \\
&=&\int_{t(x_i-\delta)}^{t(x_i)} \Phi e^{-\lambda \tau}d\tau \nonumber \\ 
&=&\frac{\Phi}{\lambda} \left( e^{-\lambda t(x_i-\delta)} - e^{-\lambda t(x_i)} \right) \label{eq:trueAsuppo}
\end{eqnarray}

Since the supported $^{210}Pb$ can be assumed to be constant, the supported activity of sample $i$ is   
\begin{eqnarray}
A_{i}^{S} &=& \int_{x_{i}-\delta}^{x_i} \rho(z) P^{S}(z)dz\nonumber \\
&=& P^{S} \rho_i. \label{eq:suppA}
\end{eqnarray}

By defining $y_i= P^T_i\rho_i$

\begin{eqnarray}
y_i\mid P^S, \Phi, \bar{t}\sim \mathcal{N} \left( A_i^{S}+\frac{\Phi}{\lambda} \left( e^{-\lambda t(x_i-\delta)} - e^{-\lambda t(x_i)} \right), (\sigma_i\rho_i)^2 \right). \label{eq:sampledis}
\end{eqnarray}
It is important to note that the activity at each sample contains not only the information of ages but also of the supported $^{210}Pb$ ($P^{S}$) and the initial supply of unsupported $^{210}Pb$ ($\Phi$) throughout the core.

To implement a Bayesian approach, prior distributions for each parameter have to be defined. \cite{Appleby2001} suggested that the supply of unsupported $^{210}Pb$ has a global mean of $50 \frac{Bq}{m^2yr}$. This can be used as prior information to obtain a prior distribution for $\Phi$. Because $\Phi$ is always positive, a gamma distribution can be considered and with this information we can define $\Phi\sim Gamma(a_\Phi,b_\Phi)$. On the other hand, since supported $^{210}Pb$ ($P^{S}$) varies much from site to site, defining an informative prior distribution for $P^S$ is in general not possible. As a consequence, data for the supported $^{210}Pb$ ($P^S$) is necessary $(y_1^{S}, y_2^{S},...,y_{n_s}^{S})$. These data can come from two different sources; $^{226}Ra$ estimates or $^{210}Pb$ measurements which reached background. A prior distribution for the $P^S$ (supported $^{210}Pb$) associated with these data is necessary. Little is known about this parameter prior to obtaining the data. We have seen cores ranging from nearly $0$ up to almost $50$ $Bq/kg$ of supported $^{210}Pb$. With this information, a gamma distribution with a mean of $20$ $Bq/kg$ and shape parameter of $\alpha_S=2$ would allow the data to contribute more to the posterior value of $P^{S}$. Lastly, to define a prior distribution for the ages an age-depth function has to be defined.

\subsection{Age-depth function}

Since sediment cores can extend back thousands of years, $^{210}Pb$ is not the only technique used to date them. $^{14}C$ (radiocarbon) is a common way to obtain age estimates for organic material. The radiocarbon community has built sophisticated chronology models, which rely on equally sophisticated age-depth functions, with the objective of reducing and better representing the uncertainty of the resulting chronology. Because we want our approach to have the flexibility to incorporate other dating information such as radiocarbon, we decided to incorporate a well-established age-depth function.

{\itshape Bacon} \citep{Blaauw2011}, which is one of the most popular chronology models for $^{14}C$ dating, assumes linear accumulation rates over segments of equal length. By using this same structure, age-depth models based on multiple isotopes could potentially be obtained. This is the age-depth model we are going to use and we discuss the general construction of the {\itshape Bacon} age-depth function next \citep[see][for details]{Blaauw2011}. The age-depth function is considered as linear over sections of equal length, causing depths to be divided into sections of equal length $c_0<c_1<c_3<...<c_K$, noting that $c_0=0$. Between these sections linear accumulation is assumed, so for section $c_i<d<c_{i+1}$ the model can be expressed as

\begin{eqnarray}
G(d,m)= \sum_{j=1}^i m_j \Delta c  + m_{i+1} (d-c_i)   , \label{eq:Bacon}
\end{eqnarray}
where $m=(m_1,m_2,...,m_k)$ are the slopes of each linear extrapolation, and
$\Delta c=c_i-c_{i+1}$ is the length of each section. 

With this structure a gamma autoregressive model is proposed for the accumulation rate of each section, $m_j=\omega m_{j+1}+(1-\omega)\alpha_j$ where $\alpha_j \sim Gamma(a_\alpha,b_\alpha)$ and $\omega\in [0,1]$ is a memory parameter which is distributed prior to the data as $\omega \sim Beta(a_\omega,b_\omega)$.

Using the above age-depth function, and (\ref{eq:sampledis}), the log-likelihood for the model takes the form
\begin{eqnarray}
\ell(\bar{y},\bar{y}^S \mid m,\omega,\Phi,P^S )&\propto  -\sum_{i=1}^n \frac{\left(y_i-\left(A_i^S+ \frac{\Phi}{\lambda}\left( e^{-\lambda G(x_{i-1},m)} - e^{-\lambda G(x_{i},m)}\right) \right)\right)^2}{2\sigma_i^2} - \sum_{j=1}^{n_s} \frac{(y_j^S-P^S)}{2\sigma_j^2}. \label{eq:lik01}
\end{eqnarray}

This likelihood contains all the parameters needed by our approach. Using the prior distributions previously mentioned, a posterior distribution $f( m,\omega,\Phi,P^S \mid \bar{y},\bar{y}^S )$ is defined, from which we intend to Monte Carlo sample the model parameters using MCMC. To allow for faster convergence of the MCMC, a limit to the chronology is considered. This chronology limit is inspired by the $^{210}Pb$ dating horizon, which is the age at which $^{210}Pb$ samples lack any measurable unsupported $^{210}Pb$.  

\subsection{Chronology limit}

The $^{210}Pb$ dating horizon was described by \cite{Appleby1998} to be 100 - 150 years, based on the available knowledge and measurement techniques at the time. The dating horizon of a given core is affected by different factors. The first of them is the equipment used to measure the samples. If certain equipment has higher precision than another, it will be able to distinguish unsupported from supported $^{210}Pb$ down to deeper samples and thus provide ages further back in time. The other factor that affects the dating horizon is the quantity of initial unsupported $^{210}Pb$, which is directly affected by the rate of supply ($\Phi$). When there is a larger initial unsupported $^{210}Pb$ it will take longer for the unsupported $^{210}Pb$ in a sample to become indistinguishable from the supported $^{210}Pb$. 

We therefore decided to set a dynamic chronology limit for our method. This limit ($t_l$) will be determined by two factors -- the rate of supply of $^{210}Pb$ to the site ($\Phi$) and the error related to the equipment used to measure the samples. For example, lets assume that the equipment used to calculate the concentration of $^{210}Pb$ in a sample has a minimum error of $0.01\frac{Bq}{kg}$. Now, assuming that the sample comes from a bog with a density ranging between $.05$ to $0.2$ $\frac{g}{cm^3}$ \citep{Chambers2011}, then once the unsupported activity in a sample reaches $A_l \simeq.1\frac{Bq}{m^2}$, it would become indistinguishable from the supported activity. This information could help us to calculate the dynamic age limit. By using equation (\ref{eq:trueAsuppo}) we have 
\begin{eqnarray}
A_l &=\int_{t_{l}}^{t_l+1} \Phi e^{-\lambda \tau}d\tau \nonumber \\
			&= \Phi e^{-\lambda t_{l}}\frac{1-e^{-\lambda}}{\lambda},
\end{eqnarray}
where $A_l$ is the minimum distinguishable unsupported activity in a sample related to the equipment's error, $\Phi$ is the supply of $^{210}Pb$ to the site and $\lambda = 0.03114$ is the decay constant and considering that $\frac{1-e^{-\lambda}}{\lambda}=0.98459$ is a constant,
\begin{eqnarray} 
 t_l &= \frac{1}{\lambda}\log\left(\frac{0.98459 \Phi}{A_l}\right) \nonumber \\
 	&\simeq \frac{1}{\lambda}\log\left(\frac{\Phi}{A_l}\right). \label{eq:t-lim}
\end{eqnarray}

It is important to note that this limit depends on the error of the equipment and on the origin of the samples, which are factors known prior to obtaining the data. Moreover, $\Phi$ is a variable of the model. This will allow the model to limit the chronology given $\Phi$.

\subsection{Implementation and MCMC}

\cite{Blaauw2011} propose the use of a self-adjusting MCMC algorithm, known as t-walk \citep{christen2010}, which will facilitate the use of these techniques to non-statisticians. The t-walk algorithm requires two initial points for all parameters ($\Phi, P^{S},  w,\alpha$) and the negative of the log posterior function which is called the energy function,
\begin{eqnarray}
U(\Phi, P^{S}, w, \alpha \mid \bar{y},\bar{y}^S)&=-\log f\left(\Phi, P^{S}, w, \alpha \mid \bar{y},\bar{y}^S \right) .
\end{eqnarray}
A program (in python 2.7) called {\itshape Plum} is used to implement this approach and to obtain a sample from the posterior distribution. {\itshape Plum} has been tested on peat and lake sediment cores, as well as on simulated data, providing reasonable results with no tuning of the parameters needed; examples of these results can be seen in sections \ref{sec:compa} and \ref{sec:simsample}. The consistency of these results, with minimal user input, show how the t-walk \citep{christen2010} was a suitable choice for this implementation.

\section{Model comparison}\label{sec:compa}

\begin{figure}[ht!]
	\centering
	\includegraphics[height=.9\linewidth]{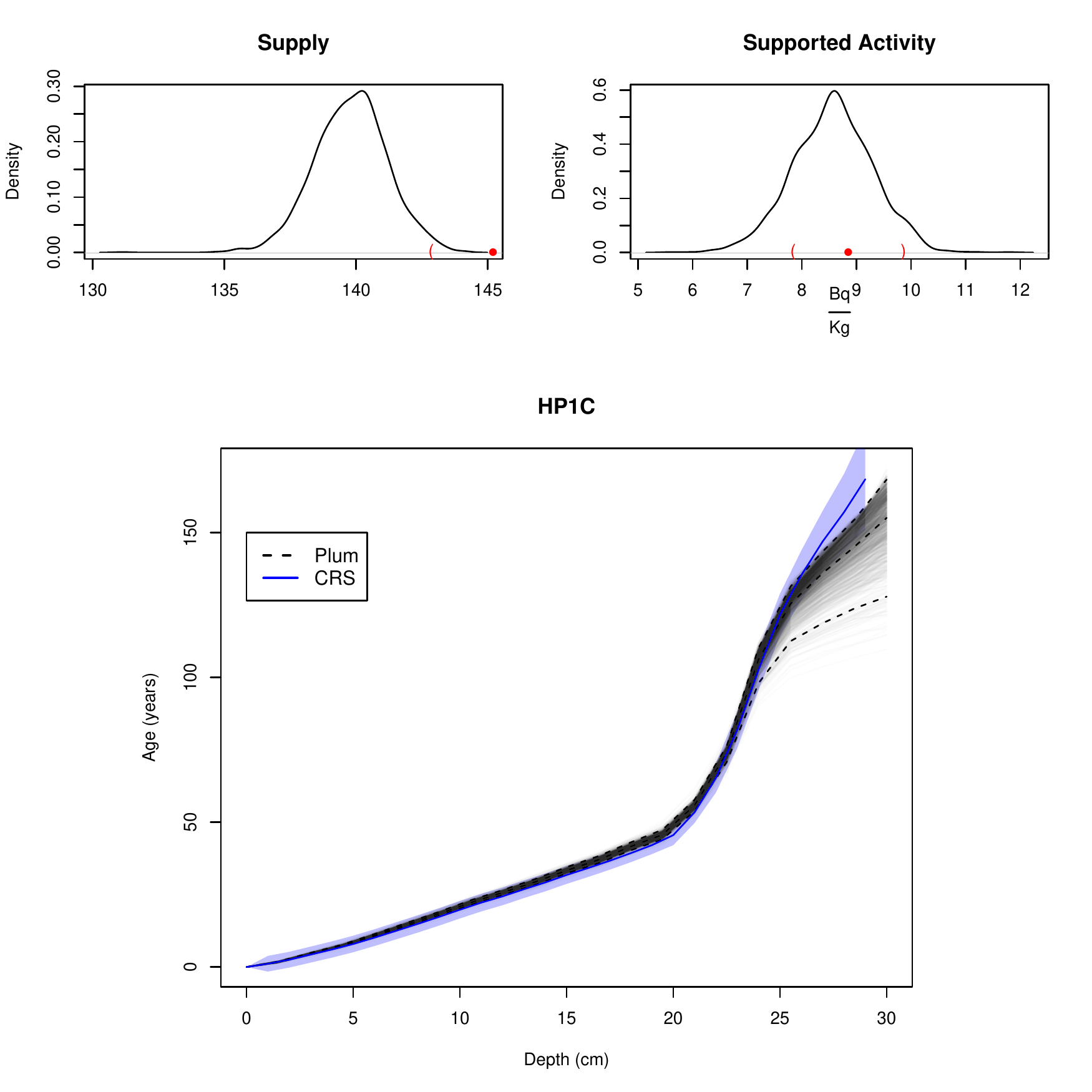}
	\caption{Comparison between the CRS \protect\citep{Appleby2001} and our model using data from HP1C. Blue curve and shadow indicate CRS mean and its corresponding 95\% range. Dashed black curves indicate mean and 95\% confidence interval for our model. Grey lines are simulations from \textit{Plum}. The top curves represent estimates of the supply of unsupported $^{210}Pb$ ($\Phi$) and supported $^{210}Pb$ ($P^S$) using the CRS model (red; dot shows the mean, parentheses show the standard deviation) and \textit{Plum} (black curve). } 
	\label{fig:San-chrono}
\end{figure}

To implement our approach to the HP1C data presented in section \ref{sec:HP1CCRS}, {\itshape Plum} was programmed to use the last 4 samples from Table \ref{Tab:DataHP1C} as estimates of the supported activity, using the rest of the samples to establish the chronology. Figure \ref{fig:San-chrono} shows the results of the CRS model in red and our approach in black and grey. From this comparison we can observe that both models agree with each other down to a depth of $25$ cm, at which point the CRS model continues at a similar slope unlike our approach which provides younger estimates. This uninterrupted growth of the CRS model can be explained by its age function which is a logarithmic approximation, invariably tends to infinity as unsupported $^{210}Pb$ reaches 0 . Even with these discrepancies both models have overlapping confidence intervals, with our approach providing a more precise chronology in the topmost part and a more conservative estimate for the deepest part of the core. 

This example shows the potential of our approach in a `well-behaved' real-world case study, but unfortunately we cannot observe the precision of this approach when confronted with more challenging data sets, such at those which did not reach background and/or with missing data. For this purpose, several simulations were created where we know the `true' chronology and can observe how our approach behaves in more challenging circumstances. 

\newpage
\section{Simulated Example}\label{sec:simsample}

To obtain simulated data, a constant supply of $^{210}Pb$ was defined as $\Phi = 150\frac{Bq}{kg}$, and by using the constant rate of supply assumption from equation (\ref{eq:const}) we have $P_0(x)r(x) = 150$. At this point, we can define $\rho(x)$ to obtain $r(x)$ by using equation (\ref{eq:drysed2}) and the age function $t(x)=  x^2/3 + x/2$.
\begin{eqnarray}
\rho(x)&=& 1.5-.05\cos\left(\frac{x}{30\pi}\right) \\
P_0(x)&=& \frac{150(\frac{2x}{3}+\frac{1}{2})}{\rho(x)}.
\end{eqnarray}

Using these functions, simulated samples at any given depth can be obtained by integrating each function between the top and bottom depths of the sample. Lastly, to simulate supported $^{210}Pb$ a constant value was added to the simulations such that $P_i= P^{S}+\int_{a}^{b}P^{U}(x)dx$, where $a$ and $b$ are the top and bottom depths of the sample. For this simulation we set the supported $^{210}Pb$ to $P^{S}=20$. To replicate the measurement errors related to the concentration of $^{210}Pb$, white noise was added such that $P_i+\epsilon$ where $P_i$ is the concentration found in sample $i$ and $\epsilon\sim \mathcal{N}\left(0,\sigma_i\right)$. This exercise provided us with the dataset in Table \ref{Tab:Data}. We use this simulated data set to test the precision of our approach in various circumstances. For this purpose, the last three sample points were designated as estimates of the supported $^{210}Pb$.

\begin{figure}[ht!]
	\centering
	\includegraphics[width=.9\linewidth]{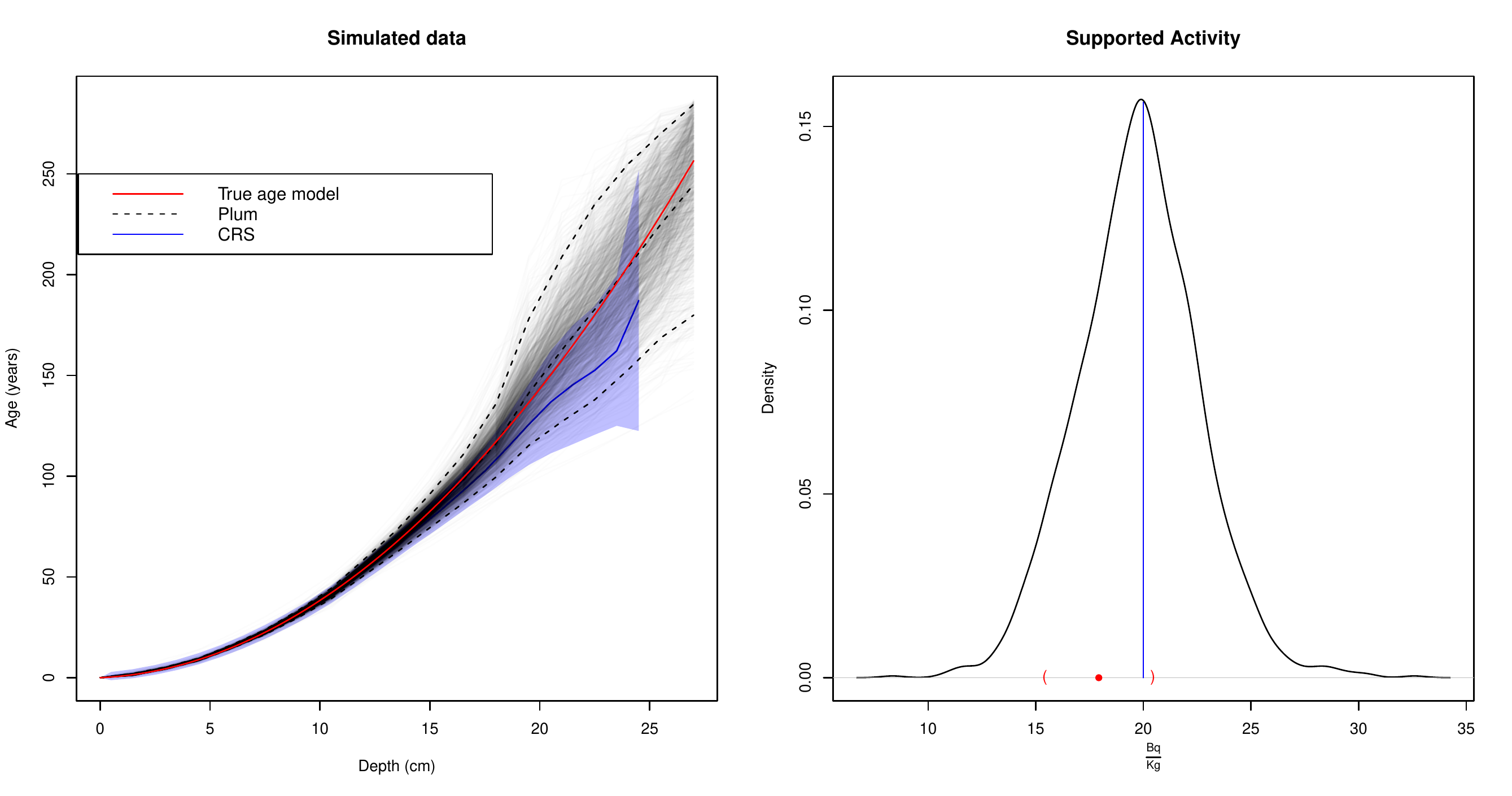}
	\caption{Comparison between the CRS \protect\citep{Appleby2001} and our model using simulated data. Blue curve and shadow indicate CRS mean and its corresponding 95\% range. Dashed black curves indicate mean and 95\% confidence interval for our model. Grey lines are simulations from \textit{Plum}. Red curve is the true age-depth model. The curve in the right represent estimates of the supported $^{210}Pb$ ($P^S$) using the CRS model (red; dot shows the mean, parentheses show the standard deviation) and \textit{Plum} (black curve). True supported $^{210}Pb$ ($P^S$) is marked by a blue line. } 
	\label{fig:modelfull}
\end{figure}

The best scenario for $^{210}Pb$ age-depth models is when every core section is measured, from the surface to where background is reached.  In this scenario any approach should reach the best results, thus providing the complete information about the decay of unsupported $^{210}Pb$. This scenario can be simulated using the complete data set from Table \ref{Tab:Data}.  Figure \ref{fig:modelfull} shows the comparison between the chronology obtained by our model and that of the CRS \cite{Appleby2001} alongside the real age function, and how both models include the true chronology in their 95\% intervals. By applying our approach to this scenario, we obtained a very accurate chronology by taking the mean of the MCMC simulations. This shows, unsurprisingly, that our model behaves quite well in the best-case scenario. On the other hand, the CRS model provides a shorter chronology, because some samples
had to be discarded from the chronology. This is a direct result from the logarithmic approximation mentioned in section \ref{sec:compa}.  In this particular case, the two bottommost samples had to be discarded since
the last sample was smaller than the mean of the three samples used to calculate the supported activity. On the other hand, CRS estimates younger ages for this example, which can be a result of the underestimated supported $^{210}Pb$ as can be observed in figure \ref{fig:modelfull}. Another feature of the CRS worth mentioning is the rapid growth of the chronology in the last sample. As previously mentioned, this rapid increase can be attributed to the logarithmic approximation the CRS uses. 

The following scenarios deal with the behaviour of our model in circumstances where there is not complete dating information. Even if we attempt to use the CRS model to provide age estimate in these scenarios, it does this by interpolating and extrapolating in the sections where there is missing data.  Applying the CRS model to these simulations would
require us to take several additional heuristic decisions with large potential impacts on the chronology (e.g., exponential or linear extrapolation to beyond and/or between the dated levels etc., see \cite{Sanchez-Cabeza2012}). Such comparisons lie outside the scope of the present work but will be explored in a future study and consequently for the next examples we only study the performance of the Plum chronology.

\begin{figure}[ht!]
	\centering
	\includegraphics[height=.9\linewidth]{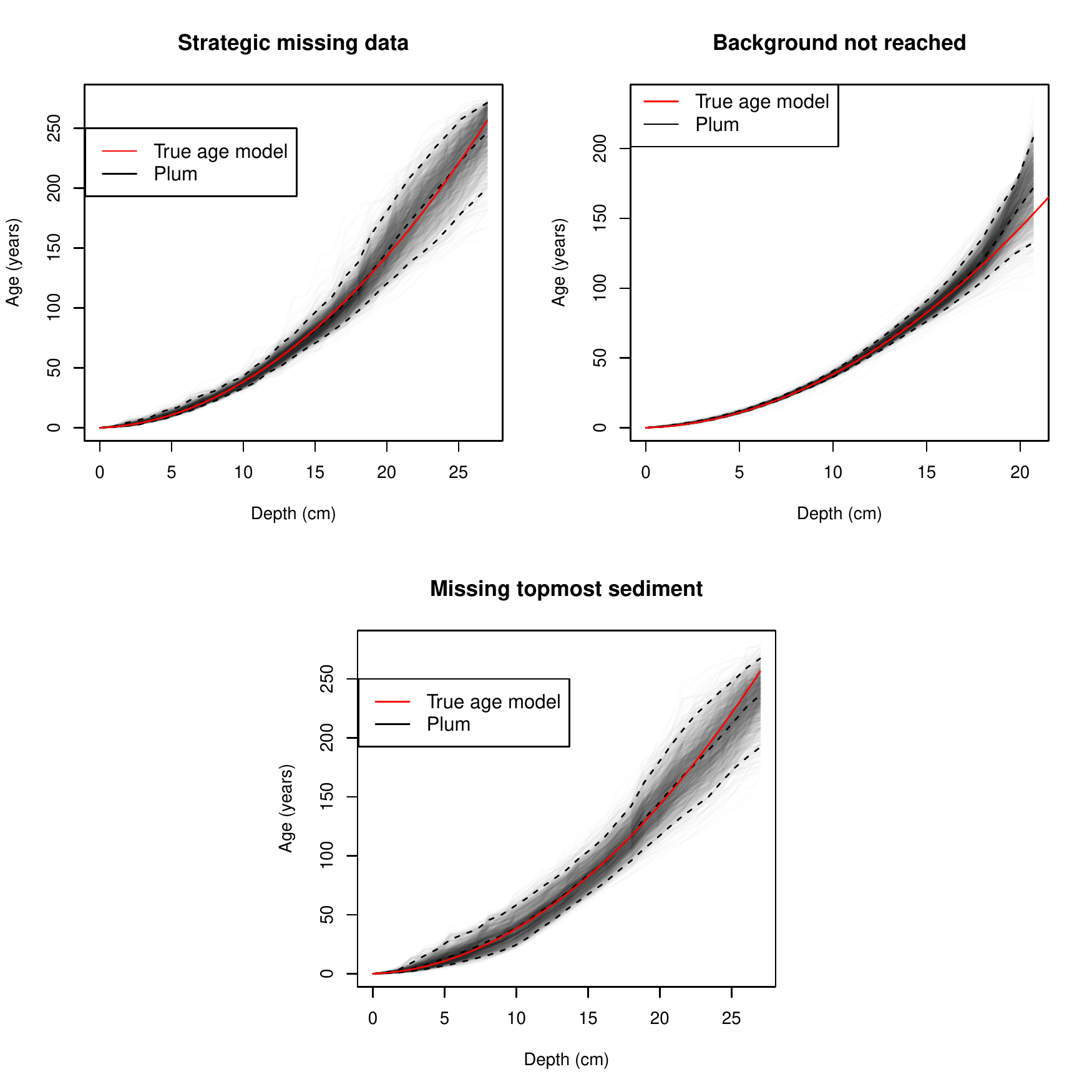}
	\caption{Bayesian analysis of simulated $^{210}Pb$ data using odd depths in the top-left, using samples with depths 1-20 in the top-right and using the samples with depths 1 and 11-27 in the bottom-centre. The red line represents the true age-depth function, grey lines are simulations from \textit{Plum}; dashed lines represent the 95\% interval and mean. }
	\label{fig:Fhalf}
\end{figure}

Sometimes researchers do not have the funds to obtain a full, continuously measured dataset for the chronology that they want to build. When this is the case, only certain strategically placed samples are measured. To simulate this scenario, only the data at odd depths was used to obtain the chronology. Figure \ref{fig:Fhalf} shows the results from this experiment. The accuracy of the model did not change as it still gives an accurate estimate of the true age model, and the precision was not greatly affected even though only half of the available data was used to calculate this chronology.

A common problem in $^{210}Pb$ dating is not reaching background. To observe the behaviour of our model in these circumstances, the bottommost seven data points were removed leaving us with a dataset which has not reached background. Figure \ref{fig:Fhalf} presents the resulting chronology compared to the true age function. The chronology seems to be accurate down to a depth of 16 cm, from which point it seems to provide older estimates. On the other hand, the model encloses the true chronology at all times even for the older ages.

The last scenario to which our approach was tested is missing topmost sediment. For this example, the data points with a depth of two to ten cm depth were removed leaving us with a data set with topmost missing data. Figure \ref{fig:Fhalf} shows the results of this experiment. Even with a third of consecutive missing data, the model is able to accurately reconstruct the true age function.

Our approach behaves well in every tested scenario, as its accuracy is not greatly affected by any of the different scenarios we introduced.

\section{Discussion}

The approach developed here presents a more robust methodology to deal with $^{210}Pb$ data. The advantage comes from a more realistic measure of uncertainties, since the ages are parameters which are inferred in the process. Moreover, dealing with missing data, which is a common problem when dealing with $^{210}Pb$ dating, becomes easier because our model does not need the whole core to be measured to obtain accurate results. Also, since the CRS model relies on a ratio, that approach requires removal of the bottommost measurement. Since our methodology does not rely on a ratio, all the samples provide information to the chronology, making longer chronologies possible. 

Because of the integration of the supported $^{210}Pb$ into the model a posterior distribution of this variable can be obtained, as well as for ages at any given depth (not just those with $^{210}Pb$ measurements) and the supply of $^{210}Pb$ to the site. Figure \ref{fig:San-chrono} shows the posterior distributions of the supported $^{210}Pb$ and the supply of unsupported $^{210}Pb$. These posterior distributions provide more realistic estimates of the uncertainty of these variables, which may be used for other studies where the main focus is not the chronology but other aspects of the site. 

Another advantage of this methodology is the fact that since the model operates within a Bayesian framework, incorporating extra information is possible without having to `double-model' by using previously modelled ages within an age-depth model. This information could come in the form of other radiometric ages, such as radiocarbon determinations. Since measurements of radiocarbon and $^{210}Pb$, given the age, are independent, the overall likelihood would consist of two parts; the likelihood from $^{210}Pb$ and from $^{14}C$.  Therefore, 
\begin{eqnarray}
\mathcal{L}\left(\Theta\right) = \mathcal{L}_{210Pb}\left(\Theta\right) \mathcal{L}_{14C}\left(\Theta\right).
\end{eqnarray}

Considering that the only link between both data is $t(x)$, by using the same age-depth function such as that from equation (\ref{eq:Bacon}), a chronology with both sources of data is possible. This becomes very important because the calibration curve \citep{Reimer2013}, which is used to correct the radiocarbon ages, is non-linear for the most recent few centuries, causing problems with interpreting radiocarbon ages. This period is partly covered by $^{210}Pb$. By combining these two methodologies,  more robust chronologies can be obtained for this important period in human and environmental history.

\bibliographystyle{ECA_jasa.bst}
\bibliography{Plum_Bib.bib}
\newpage

\section{Data}

\begin{table}[ht!] \centering 
	\begin{tabular}{cccc|cccc} 
		\hline 
		Depth  & $^ {210}Pb$  & $\sigma$($^ {210}Pb$)  & Density ($\rho$) & Depth  & $^ {210}Pb$  & $\sigma$($^ {210}Pb$)  & Density ($\rho$)   \\ \hline
		$cm$  & $Bq/kg$  & $\sigma(Bq/kg)$ & $g/cm^2$ & $cm$  & $Bq/kg$  & $\sigma(Bq/kg)$ & $g/cm^2$ \\ \hline 
		$1$ & $371.730$ & $11.900$ & $0.045$ &$18$ & $279.320$ & $11.140$ & $0.045$ \\ 
		$2$ & $456.390$ & $15.080$ & $0.047$ &$19$ & $243.820$ & $9.940$ & $0.045$ \\ 
		$3$ & $454.240$ & $17.110$ & $0.051$ &$20$ & $246.750$ & $9.170$ & $0.054$ \\  
		$4$ & $449.640$ & $14.430$ & $0.049$ &$21$ & $351.680$ & $13.100$ & $0.086$ \\ 
		$5$ & $479.040$ & $16.440$ & $0.049$ &$22$ & $281.280$ & $11.380$ & $0.089$ \\  
		$6$ & $490.970$ & $16.750$ & $0.051$ &$23$ & $235.300$ & $12.720$ & $0.099$ \\  
		$7$ & $482.120$ & $16.780$ & $0.050$ &$24$ & $192.820$ & $7.240$ & $0.085$ \\ 
		$8$ & $486.880$ & $15.200$ & $0.047$ &$25$ & $94.280$ & $4.740$ & $0.066$ \\ 
		$9$ & $431.580$ & $14.830$ & $0.048$ &$26$ & $50.550$ & $3.410$ & $0.060$ \\  
		$10$ & $422.750$ & $16.210$ & $0.049$ &$27$ & $36.080$ & $2.260$ & $0.062$ \\ 
		$11$ & $315.310$ & $13.030$ & $0.052$ &$28$ & $28.710$ & $2.100$ & $0.055$ \\ 
		$12$ & $349.770$ & $15.220$ & $0.047$ &$29$ & $24.680$ & $1.760$ & $0.059$ \\  
		$13$ & $301.740$ & $13.450$ & $0.051$ &$35$ & $11.040$ & $1.270$ & $0.356$ \\  
		$14$ & $284.410$ & $10.020$ & $0.050$ &$40$ & $6.240$ & $1.010$ & $0.414$ \\ 
		$15$ & $280.580$ & $11.620$ & $0.053$ &$45$ & $10.150$ & $1.310$ & $0.347$ \\ 
		$16$ & $250.170$ & $9.760$ & $0.048$  &	$50$ & $7.960$ & $1.600$ & $0.352$ \\ 
		$17$ & $267.740$ & $12.950$ & $0.048$ &	$ $ & $ $ & $ $ & $ $ \\
	\end{tabular} 
	\caption{HP1C dataset. This table presents the necessary information to replicate the results from the CRS model as well as from our approach.} 
	\label{Tab:DataHP1C} 
\end{table}

\begin{table}[ht!]
	\centering
	\begin{tabular}{cccc|cccc}
		\hline
		Depth  & $^ {210}Pb$ ($P^T$) & $\sigma$($^ {210}Pb$) & Density ($\rho$) & Depth  & $^ {210}Pb$ ($P^T$) & $\sigma$($^ {210}Pb$) & Density ($\rho$)\\ 
		\hline
		$cm$  & $Bq/kg$  & $\sigma(Bq/kg)$ & $g/cm^2$ & $cm$  & $Bq/kg$  & $\sigma(Bq/kg)$ & $g/cm^2$ \\ \hline 
			$1$ & $102.897$ & $10$ & $0.145$ & $16$ & $80.845$ & $7$ & $0.150$  \\ 
			$2$ & $180.761$ & $9$ & $0.145$ & $17$ & $64.024$ & $7$ & $0.151$    \\ 
			$3$ & $220.507$ & $9$ & $0.145$ & $18$ & $48.792$ & $7$ & $0.151$    \\ 
			$4$ & $268.669$ & $9$ & $0.145$ & $19$ & $54.076$ & $7$ & $0.152$    \\ 
			$5$ & $285.026$ & $9$ & $0.146$ & $20$ & $37.109$ & $7$ & $0.152$    \\ 
			$6$ & $311.949$ & $9$ & $0.146$ & $21$ & $36.640$ & $7$ & $0.153$    \\ 
			$7$ & $298.226$ & $9$ & $0.146$ & $22$ & $28.602$ & $7$ & $0.153$    \\ 
			$8$ & $302.736$ & $9$ & $0.146$ & $23$ & $22.180$ & $6$ & $0.154$    \\ 
			$9$ & $262.598$ & $8$ & $0.147$ & $24$ & $29.342$ & $6$ & $0.154$  \\ 
			$10$ & $251.080$ & $8$ & $0.147$ &$25$ & $28.723$ & $6$ & $0.154$  \\ 
			$11$ & $221.818$ & $8$ & $0.148$ &$26$ & $26.123$ & $6$ & $0.154$  \\ 
			$12$ & $199.937$ & $8$ & $0.148$ &$27$ & $17.803$ & $6$ & $0.155$  \\ 
			$13$ & $161.476$ & $8$ & $0.149$ &$28$ & $23.349$ & $6$ & $0.155$  \\ 
			$14$ & $132.268$ & $8$ & $0.149$ &$29$ & $13.607$ & $6$ & $0.155$  \\ 
			$15$ & $112.069$ & $8$ & $0.150$ &$30$ & $16.825$ & $5$ & $0.155$  \\ 
		\hline
	\end{tabular} 	
	\caption{Simulated dataset. This table presents the necessary information to replicate \protect{\itshape Plum}'s results as well as those of the CRS model \protect\citep{Appleby2001}.}
	\label{Tab:Data}
\end{table}

\end{document}